\documentclass[]{spie}
\usepackage[]{graphicx}
\usepackage[]{rotating}
\usepackage[]{aas_macros}
\usepackage{amsmath}
\usepackage{tikz}

\title{A six-apertures discrete beam combiners for J-band interferometry}

 \author{Ettore Pedretti\supit{a},  Simone Piacentini\supit{b}, Giacomo Corrielli\supit{c}, Roberto Osellame\supit{cb}, and Stefano Minardi\supit{a}
\skiplinehalf
\supit{a}Leibniz-Institut f{\"u}r Astrophysik Potsdam (AIP), An der Sternwarte 16, 14482 Potsdam, Deutschland\\
\supit{b}Dipartimento di Fisica – Politecnico di Milano, Piazza L. da Vinci 32 – 20133 Milano, Italia\\
\supit{c}Istituto di Fotonica e Nanotecnologie (IFN) – CNR, Piazza L. da Vinci 32 – 20133 Milano, Italia\\
}
\authorinfo{ 
Further author information: (Send correspondence to E.P.)\\
E.P: E--mail:epedretti@aip.de}

\usepackage[displaymath,textmath,sections,graphics,floats]{preview}
     \PreviewEnvironment{enumerate sections}

\newcommand{\twoobjects}[2]{%
  \leavevmode\vbox{\hbox{#1}\nointerlineskip\hbox{#2}}%
}

\begin{document}

\maketitle 
\def\textpm{$\pm$}
\def\textdegree{$^o$}

%%%%%%%%%%%%%%%%%%%%%%%%%%%%%%%%%%%%%%%%%%%%%%%%%%%%%%%%%%%%% 
%>>>> Include a list of keywords after the abstract 

\begin{abstract} 
The astronomical J-band (1.25 micrometres) is a relatively untapped wave-band in long-baseline infrared interferometry. It allows access to the photosphere in giant and super-giant stars relatively free from opacities of molecular bands. The J-band can potentially be used for imaging spots in the 1350 nm ionised iron line on slowly rotating magnetically-active stars through spectro-interferometry. In addition, the access to the 1080 nanometres He I line may probe outflows and funnel-flows in T-Tauri stars and allow the study of the star-disk interaction. 

We present the progress in the development of a six-inputs, J-band interferometric beam combiner based on the discrete beam combiner (DBC) concept. DBCs are periodic arrays of evanescent coupled waveguides which can be used to retrieve simultaneously the complex visibility of every baseline from a multi-aperture interferometer. Existing, planned or future interferometric facilities combine or will combine six or more telescopes at the time, thus increasing the snapshot uv coverage from the interferometric measurements. A better uv coverage will consequently enhance the accuracy of the image reconstruction. DBCs are part of the wider project “Integrated astrophotonics” that aims to validates photonic technologies for utilisation in astronomy.

Before manufacturing the component we performed extensive numerical simulations with a coupled modes model of the DBC to identify the best input configuration and array length. The 41 waveguides were arranged on a zig-zag array that allows a simple optical setup for dispersing the light at the output of the waveguides.

The component we are currently developing is manufactured in borosilicate glass using the technique of multi-pass %... To Ettore: I think it is single pass. Check with Roberto
ultrafast laser inscription (ULI), using a mode-locked Yb:KYW laser at the wavelength of 1030 nm, pulse duration of 300 fs and repetition rate of 1 MHz. After annealing, the written components showed a propagation loss less than 0.3 dB/cm and a negligible birefringence at a wavelength of 1310 nm, which makes the components suitable for un-polarized light operation. A single mode fiber-to-component insertion loss of 0.9 dB was measured. Work is currently in progress to characterize the components in spectro-interferometric mode with white light covering the J-band spectrum.

\end{abstract}

\keywords{Stellar Interferometry, Long--Baseline Interferometry,  Astrophotonics, Beam Combiners, Fibre Optics.}
\section{Introduction}

Long--baseline infrared interferometry has undoubtly benefited from the development of photonics components for the telecommunication industry. The use of integrated optics components in stellar interferometry happened before the term ``Astrophotonics'' was invented to describe %integrated--optics.. Laser Guide Stars and the like are also Astrophotonic without being IO 
photonic components applied to astronomical instrumentation. Although multi--mode pupil--plane, coaxial beam combiners in bulk optics were used in the early interferometers\cite{1975ApJ...196L..71L} the issues of complexity, calibration and stability prevented the use of bulk--optics beam combiners for interferometers composed of more than two telescopes. The difficulty of combining more than two telescopes and the lack of sensitive detectors delayed significantly the imaging of complex object with long--baseline stellar interferometry.

With the advent of fibre optics beam combiners\cite{1992ESOC...39..731C,1997ioai.book..115C,2003SPIE.4838..280C,1999ASPC..194..344R}, that used fibre couplers instead of beam splitters and beam combiners, the era of precision interferometry was born. Fibre couplers allowed the acquisition of high accuracy squared visibilities on a single telescope baseline and the measurement of accurate stellar diameters. Further advance came with the development of planar integrated optics components, in particular the IONIC\cite{1999ASPC..194..344R,2000SPIE.4006.1042R,2001PhDT.........3H}. The IONIC--3T was tested and deployed\cite{2003SPIE.4838.1099B} at the infrared optical telescope array (IOTA)\cite{1988ESOC...29..939C, 1990SPIE.1237..145T, 1995AJ....109..378D}. Together with a sentitive infrared camera\cite{2004PASP..116..377P}, the first operational 3--telescope fringe tracker\cite{2005ApOpt..44.5173P} and a low resolution spectrograph\cite{2003SPIE.4838.1225R,2008SPIE.7013E..2VP} IOTA was the first long--baseline interferometer to produce science results using ``Astrophotonics'' components\cite{2004ApJ...602L..57M,2006ApJ...645L..77M,2006ApJ...647..444M,2006ApJ...647L.127M,2006ApJ...652..650R,2007ApJ...659..626Z,2009MNRAS.397..325P,2009MNRAS.398.1309T}. %the first astrophotonic instrument producing astronomical data was MEDUSA, (1980)

The limitations of planar integrated optics became evident when combining four or more telescopes\cite{2006SPIE.6268E..2DB, 2009A&A...498..601B} due to increase in complexity when using couplers for beam combination. The idea of utilising three--dimensional structures for beam combination gained traction when ultrafast laser inscription (ULI) for astronomy became a reality\cite{2009OExpr..17.1963T,Rodenas:2012, Minardi:2012, Jovanovic:2012}. The possibility of avoiding the crossing over of couplers waveguides simplified the complexity of the beam combiners and reduced the optical crosstalk in the component. Unfortunately the low confinement of light in waveguides, intrinsic in ULI technology low--contrast waveguides, caused losses of light and low throughput from the first prototypes\cite{2017A&A...602A..66T}.
\begin{figure}[t]
\centering
\includegraphics[width=0.6\textwidth]{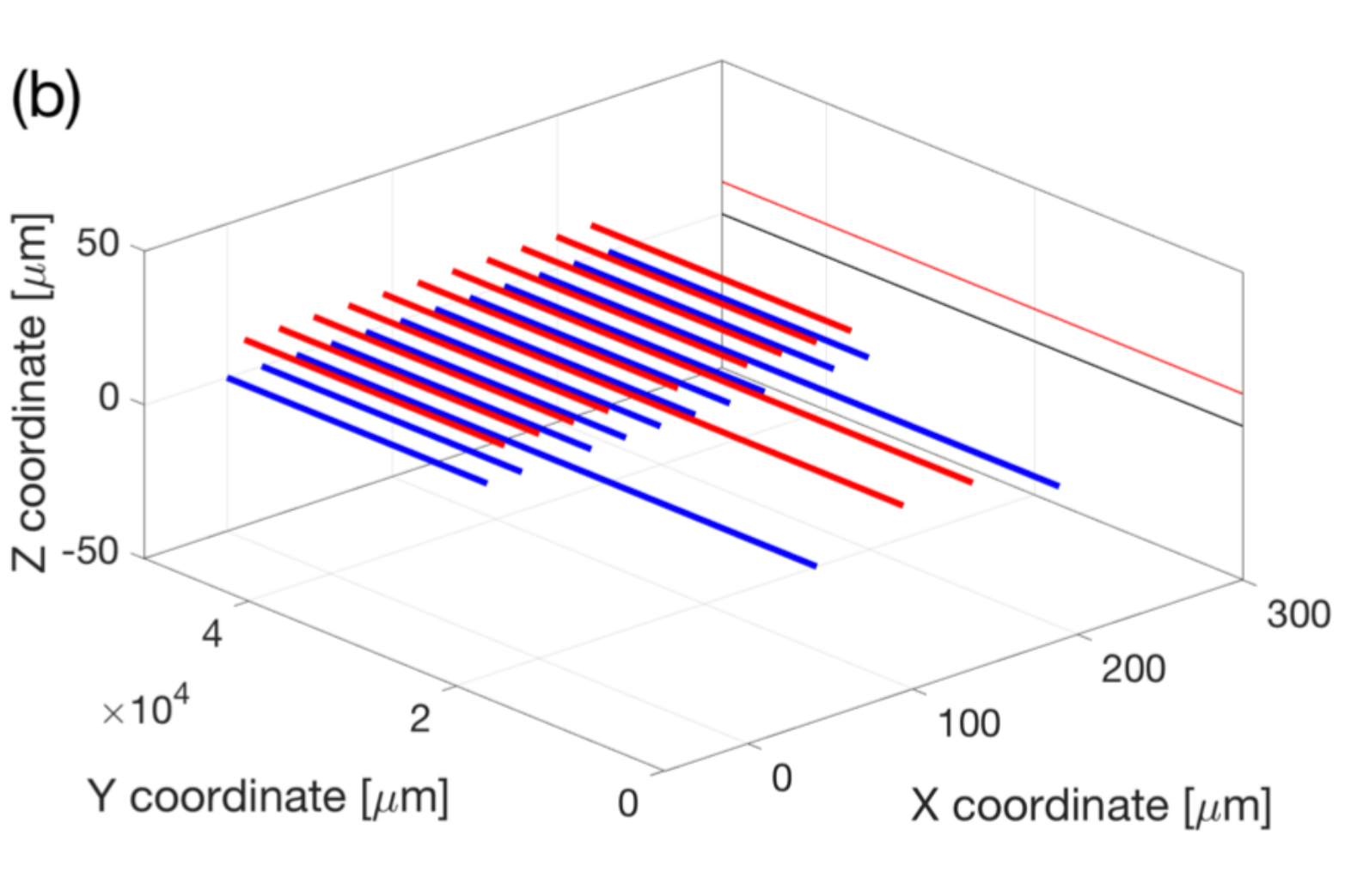} 
\caption{Example of a 4--telescope beam combiner formed from evanescent--coupled waveguides. Four input site reach the input of the component while all the output waveguides reach the output of the component.}
\label{DBCex}
\end{figure}

The discrete beam combiner (DBC) concept overcomes some of the limitations of low--throughput beam combiners written with ULI technology. The theory of the DBC is extensively described in several publications\cite{2010OptL...35.3009M,Minardi:2012,2012MNRAS.422.2656M,2013ApOpt..52.4556S,2014SPIE.9146E..1DM,2016SPIE.9907E..33E,2017OExpr..2519262D}. Briefly in the DBC concept a two-dimensional lattice of parallel waveguides is written to a glass substrate using ULI. Evanescent coupling among the waveguides allows the interferometric mixing of the \(N\) input fields at the \(M>N \times N\) outputs of the lattice, from which the complex visibilities can be retrieved from a measurement of the power at each output site. Figure~\ref{DBCex} shows an example for a 4--telescope zig-zag DBC geometry of coupled waveguide lattice.
A DBC can measure the mutual coherence of the input field or the ``complex visibility'' without the necessity of scanning the optical path that was necessary in coaxial beam combiner or the early integrated optics beam combiners like IONIC--3T. This is not dissimilar from a ``classic'' image--plane beam combiner where the intensity in the focal plane is the linear combination from the input of all the telescopes beselines. In a an image plane beam combiner the telescope baselines are spatially encoded in the interference fringe patterns recorded by a camera and the complex visibilities for each baselines can be extracted by a simple Fourier transform or by deriving the pixel to visibility matrix\cite{2007A&A...464...29T} (P2VM). In the DBC the intensity of the output waveguides can also be transformed to a complex visibilities by multiplying the outputs by the P2VM that is obtained by inverting the visibility to pixel matrix\cite{2007A&A...464...29T}  (V2PM). Explanation of how to obtain the V2PM for an image--plane beam combiner\cite{2007A&A...464...29T}, a planar integrated optics beam combiner\cite{2008SPIE.7013E..16L} and a DBC\cite{2013ApOpt..52.4556S} are given in the references.

We have manufactured a six--telescope DBC working in the astronomical J band (1.25 \(\mu m\)) using borosilicate glass and ultra--fast laser inscription. This is to our knowledge the first integrated optics beam combiner working in the J band ever manufactured.
The 1.25~\(\mu m\) astronomical J-band is relatively untapped waveband for long baseline infrared interferometry. The J band  allows access to the photosphere of giant and supergiant stars relatively free from opacities of molecular bands allowing the characterisation of large--scale convection cells on giant and supergiant stars\cite{2018Natur.553..310P}. It will enable high accuracy diameter measurement with higher resolution than the commonly used astronomical H-band. Iron (Fe ii): 1183nm 1291nm in stellar spots\cite{}. The He 1083 nm line to probe inflow and outflow in T--Tauri stars\cite{2008ApJ...687.1117F,2011MNRAS.416.2623K}.

\section{\label{subsec:DBC-J}Six telescope discrete beam combiner for the J band}
\begin{figure}[t]
\begin{center}
\twoobjects 
{\includegraphics[scale=.6]{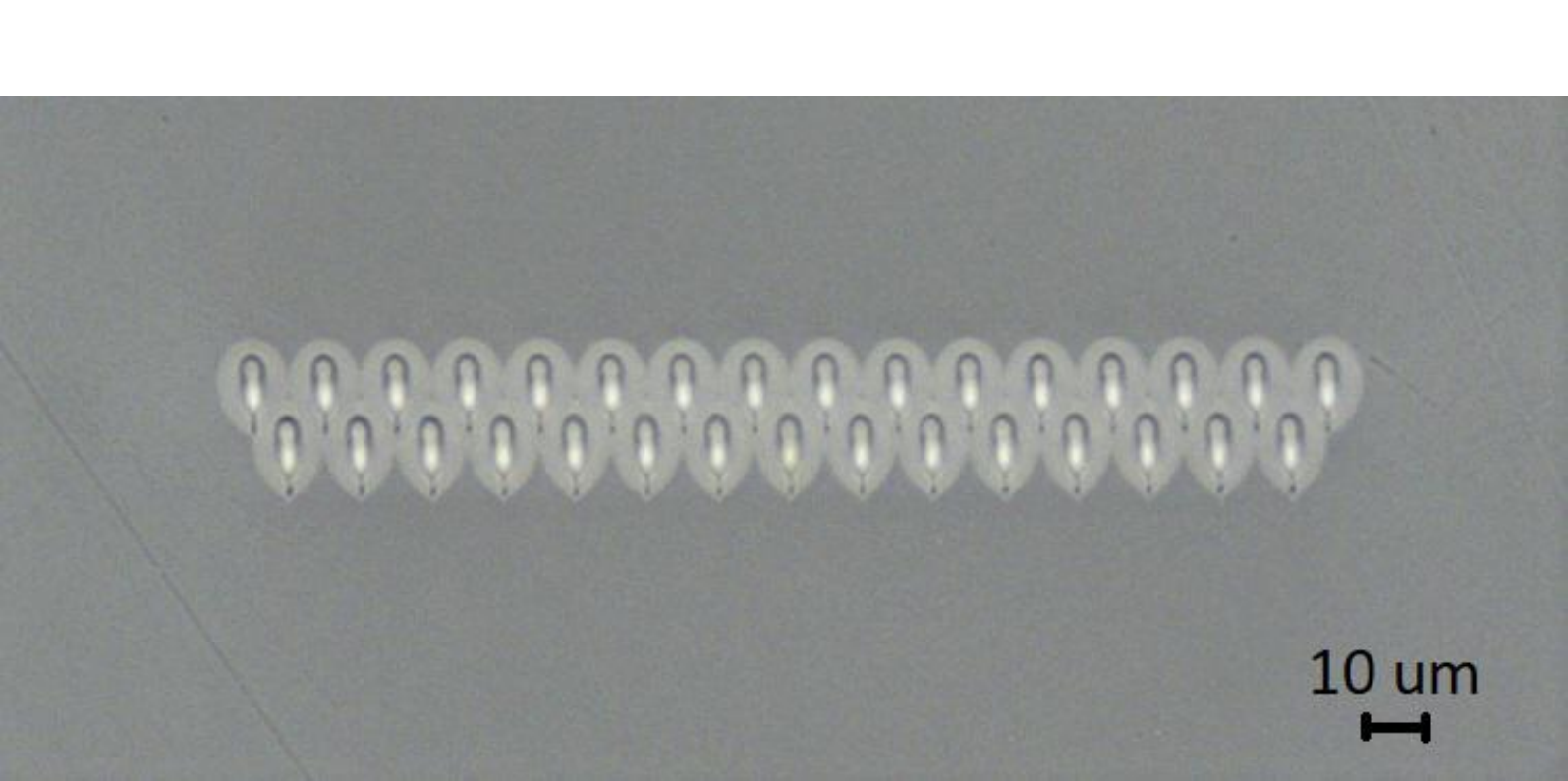}}
{\includegraphics[scale=.6]{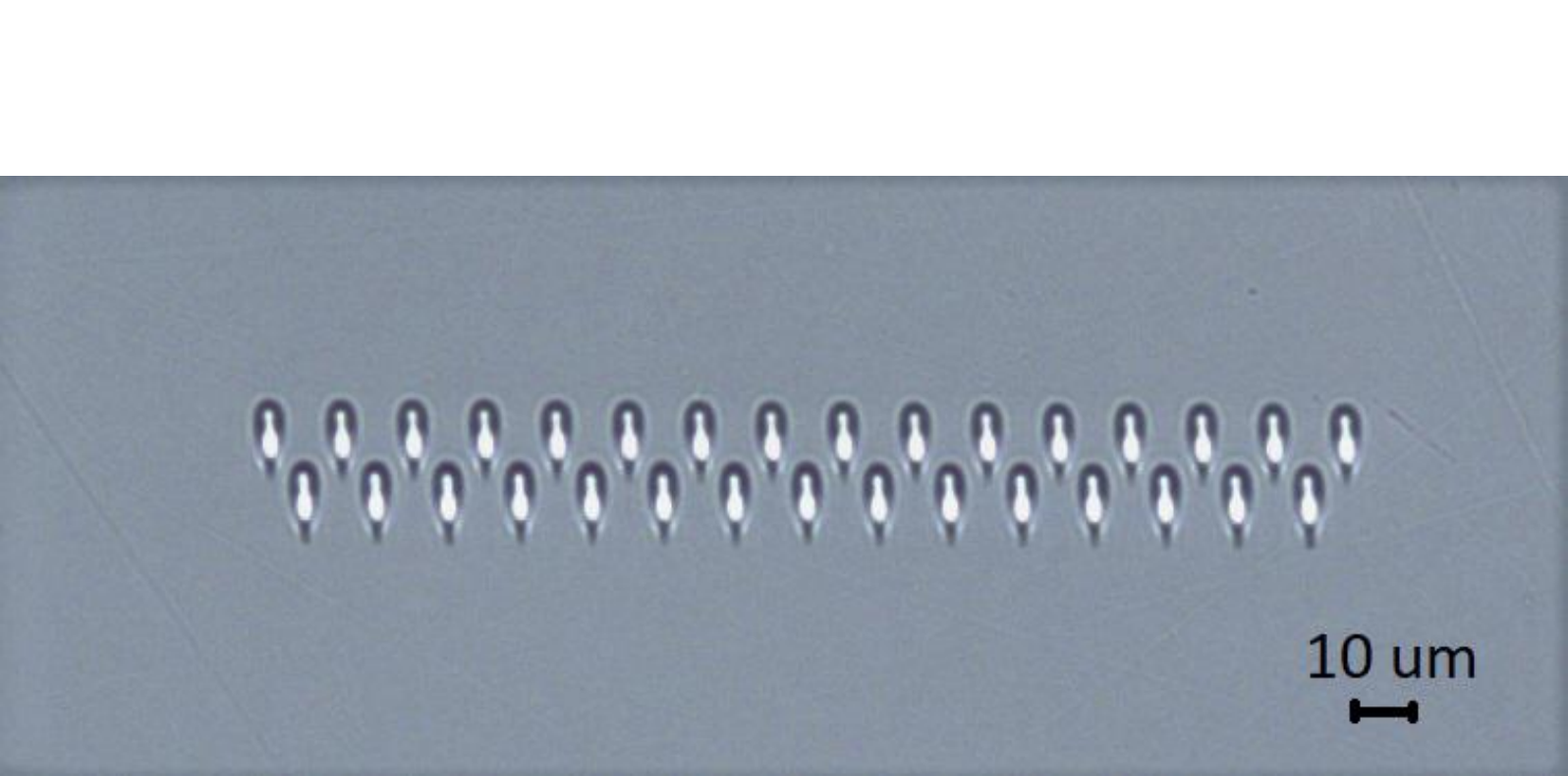}}
\end{center}
\caption{\label{fig_anneal}The component before annealing (top) and after annealing (bottom).}
\end{figure}
%A 6--telescope discrete beam combiners was manufactured for the first time for the astronomical J-band. 
The 6-telescope combiner was written in boro--alumino--silicate glass with the technique of multi--pass ultra--fast laser inscription, using a mode-locked Yb:KYW laser at the wavelength of 1030 nm, pulse duration of 300 fs and repetition rate of 1 MHz. In order to get single-mode waveguides with low losses at the wavelength of interest, we used the multi--scan technique, consisting in writing a waveguide multiple times, to increase the refractive index change. %Check with Roberto if it is single or multi-pass. I think they wrote in single pass at high power and then annealed the sample

To find the optimal writing parameters, we fabricated waveguides with different powers, speeds and number of scans, and characterised them with a 1310 nm laser diode. To study the planar coupling, we fabricated a set of planar arrays of N~=~31 waveguides with different interaction lengths (0.5~cm and 1.5~cm) and distances (from 10~\(\mu\)m to 17~\(\mu\)m, with a spacing of 1~\(\mu\)m). The arrays were then characterised by injecting in the central waveguide horizontally or vertically polarized light with a wavelength of 1310~nm. The outputs, collected with a Vidicon camera, were compared with numerical simulations in order to estimate the coupling coefficient.

After the study of the planar coupling, new devices were fabricated to investigate the dependence of the diagonal coupling on the angle between the waveguides. Indeed, a difference in the coupling strength for waveguides in and out of plane may arise from a non perfect circular mode profile\cite{2003JOSAB..20.1559O}. The geometry of the circuit is shown in Figure~\ref{fig_anneal}: two planes of waveguides with the same coupling distance are put close to each other in order to have a diagonal interaction between them (all distances between neighbouring waveguides are equal). While the central waveguide in the lower plane starts at the beginning of the sample and can be therefore coupled with light for the characterization of the device, the other ones start in the sample, so they can guide light only by power exchange with the central waveguide.

After fabrication, the waveguides were subjected to an annealing process, which reduces their birefringence by releasing laser induced stress. After annealing, (see Figure~\ref{fig_anneal}) the written components showed a propagation loss less than 0.3 dB/cm and a negligible birefringence at a wavelength of 1310 nm, which makes the components suitable for un-polarized light operation. A single mode fibre-to-component insertion loss of 0.9~dB was measured. Work is currently in progress to characterize the components in spectro-interferometric mode with white light covering the J-band spectrum. Many devices were fabricated, each one with a specific interwaveguide distance, ranging from 12~\(\mu\)m to 18~\(\mu\)m, and an array length from 4~mm to 16~mm depending on the coupling coefficient estimated in the previous measurements (in order to match the device with the design). The relative values of the diagonal and horizontal coupling coefficient was estimated by fitting the output distribution with a two-parameter coupled modes model. 

After this calibration, we fabricated a 6-telescope DBC  featuring 41 waveguides.
%DBC, to study in detail its geometry and array uniformity in a longer circuit. 
In particular, we adopted the scheme in Figure~\ref{XYscan} (bottom), which was selected as the best among all possible input configurations. Waveguides 4, 10, 16, 19, 31 and 38 start from the beginning of the sample, therefore can be coupled with light, while the other waveguides start in the glass. We employed a inter--waveguide separation of 16 \(\mu\)m and an array length length of 4 cm, which corresponds to the optimal design.
\begin{figure}[t]
\centering 
\includegraphics[width=0.8\textwidth]{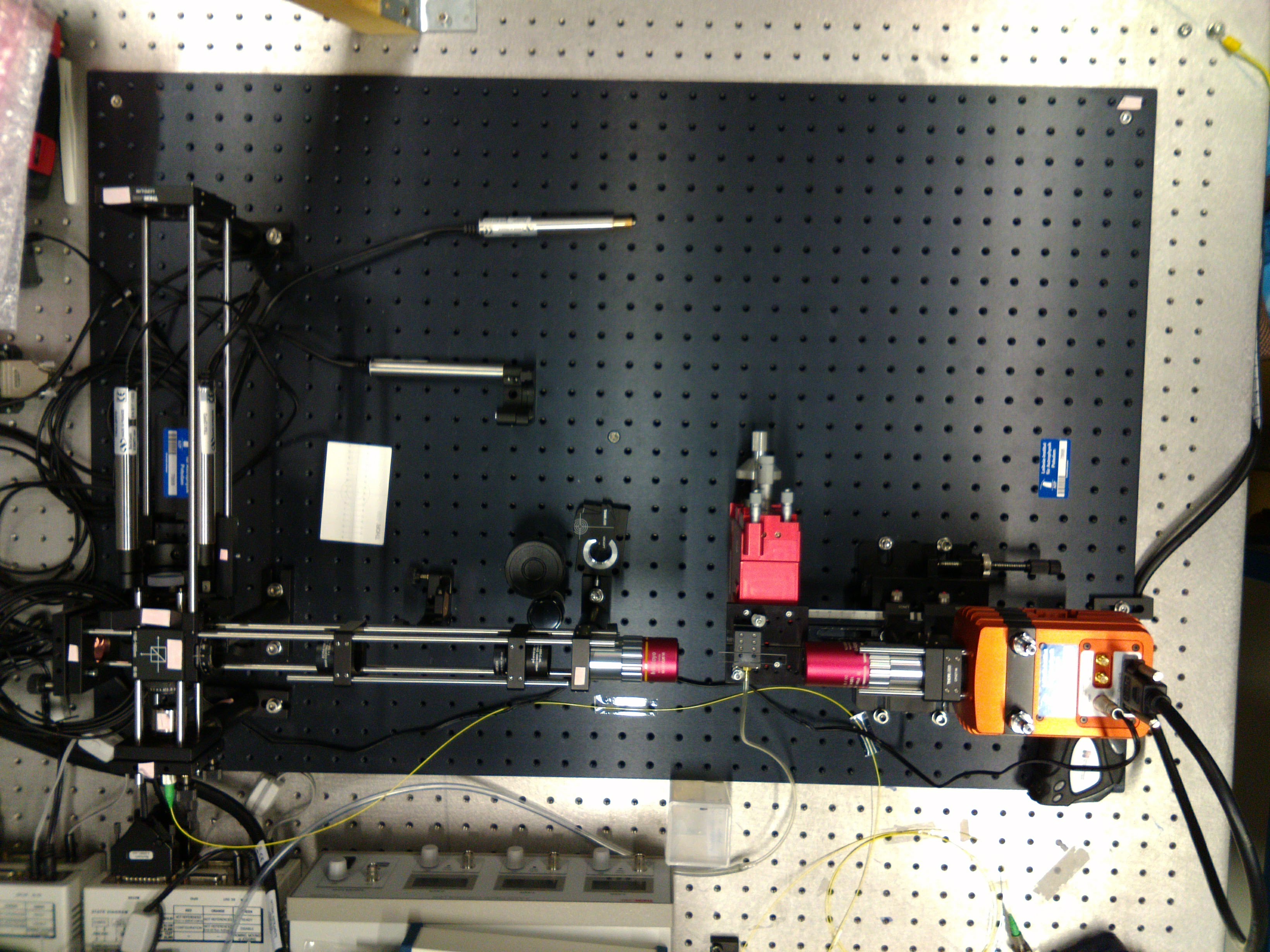} 
\caption{The J-band and H-band DBC characterization bench. The picture shows the current two beams setup at innoFSPEC in Potsdam.}
\label{JH-testbench}
\end{figure}

The interferometric test of the manufactured components usually
requires a lengthy calibration of the V2PM for all 15 input combinations.
%Since the ULI manufacturing process requires several steps of calibration 
To speed up the procedure, we set-up an automated test facility that permits to characterise single waveguides and beam combiners of 6--telescope complexity without the need of manual steering of the beams in order to find the input sites. The combiners will then be tested with a tunable laser ranging from 1250 to 1350 nm  in order to characterise the beam combiner across the whole bandwidth. 

\section{Interferometry testbench}
 The beam combiners were characterised on our automated testbench at the innoFSPEC in Potsdam. A picture of the testbench is shown in Figure~\ref{JH-testbench}. We used a Michelson interferometer capable of injecting two beams in the component through a direct and a delayed path. The delayed path uses a USB--controlled delay line connected to a computer to change the optical path difference (OPD). The chromatic response of the component can be characterised using two separate tunable laser sources: one at 1.3~\(\mu m\) for the J--band and one at 1.5~\(\mu m\) for the H--band. 
\begin{figure}[t]
\twoobjects 
 {\includegraphics[width=1\textwidth]{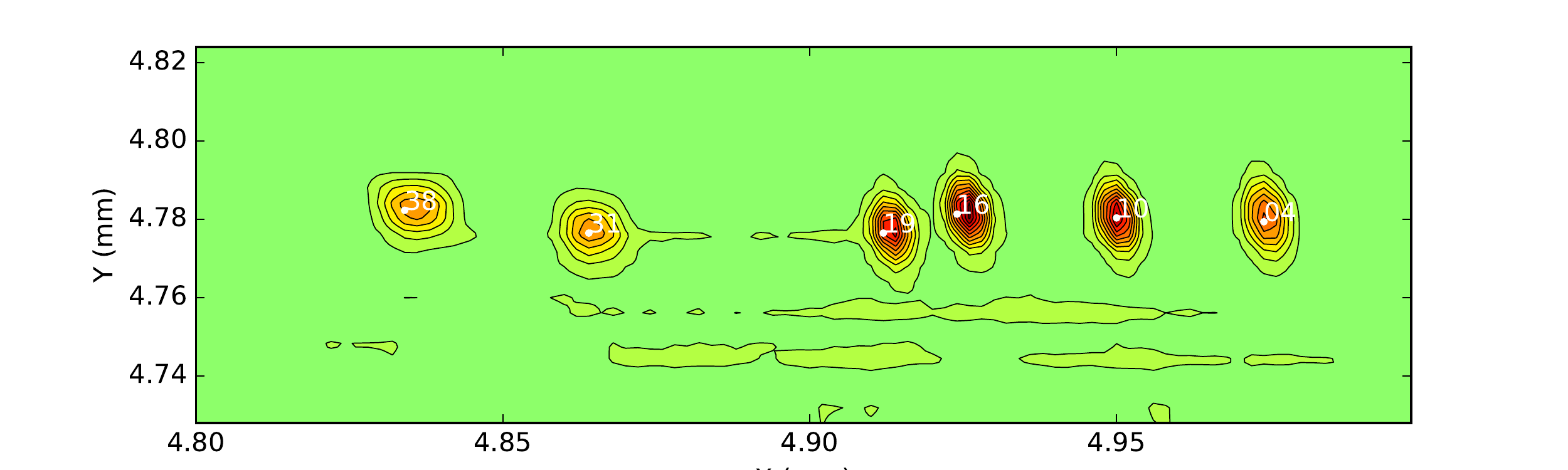}}
 {\includegraphics[width=1\textwidth]{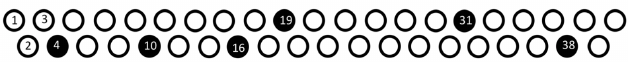}}
\caption{Top shows a map obtained by scanning in the XY direction one beam at the input of a 6--input DBC and recording the summed outputs of the components in function of the XY position. The map is necessary for identifying the position of the inputs of the component. Once the position of the inputs is known two beams can be injected and switched  rapidly to form baselines. Bottom shows a schematic of the outputs of the 6T component. The outputs in dark colour are also the site where the input beams are injected and corresponds to the positions found on the map.}
\label{XYscan}
\end{figure}

 The beam from the tunable source is split in two using a beam splitter and directed to two kinematic mounts. The mounts can be steered using stepper motors controlled by a computer through a serial link and are able to  address the different inputs of the component. Images are recorded by a fast infrared camera connected to the same computer controlling the stepper motors and the delay line. The camera is connected to the computer using a camera-link interface wired to a frame-grabber card. The a map of the inputs of the components are shown on Figure~\ref{XYscan} top.

\begin{figure}[p]
 \centering
 \includegraphics[trim={3cm 0 2.8cm 0},clip,width=1.0\textwidth]{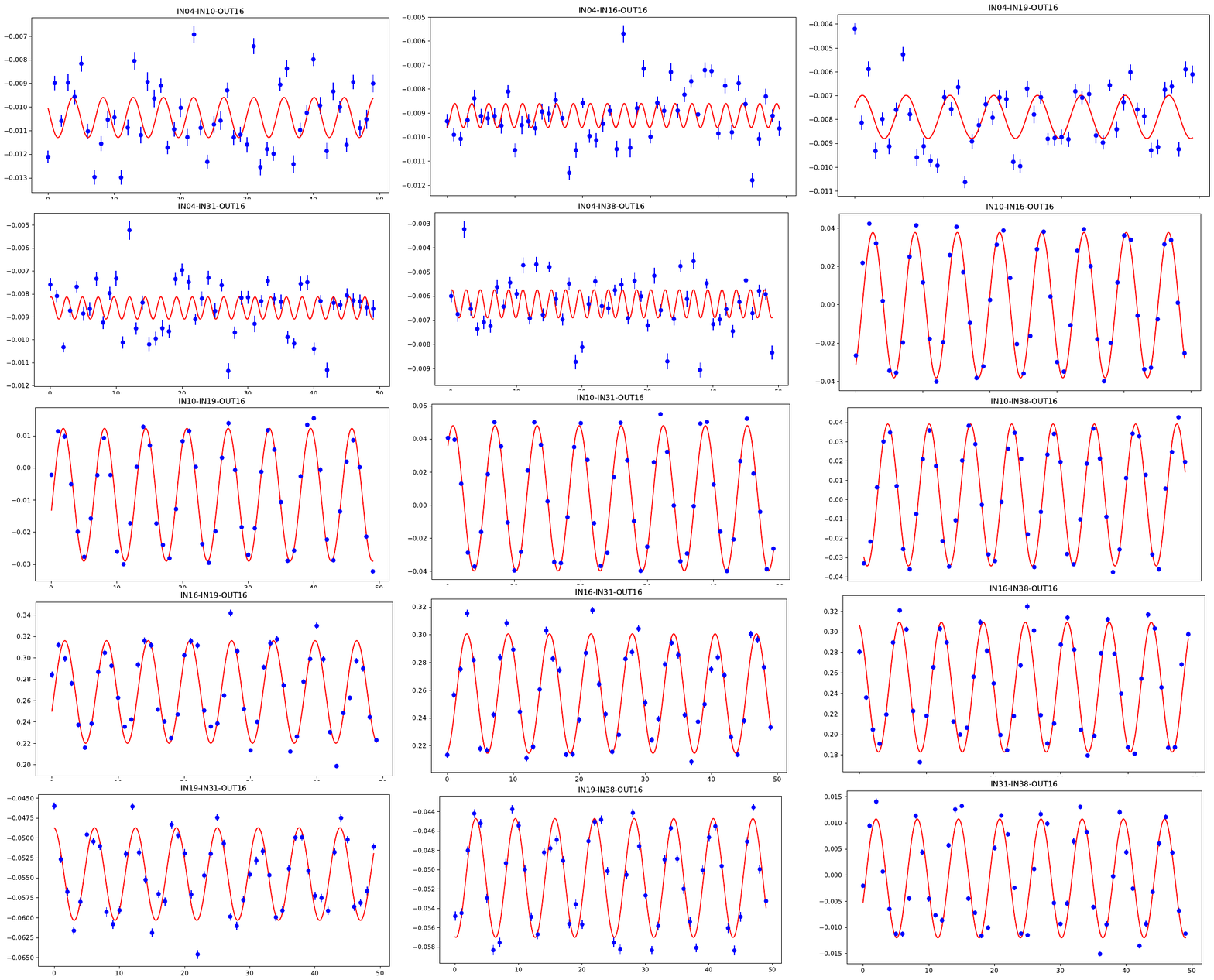}
\caption{Fringe fit for all baselines and for output 16. Five fringes have too low signal--to--noise to achieve a successful fit.}
\label{fringes}
\end{figure}

 \begin{figure}[t]
\twoobjects 
 {\includegraphics[trim=5 5 5 5,clip,width=1.0\textwidth]{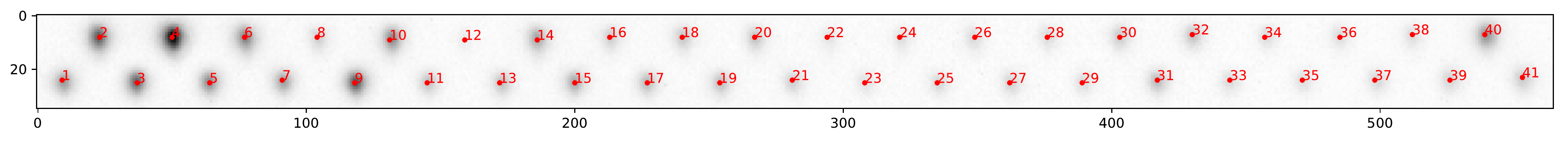}}
 {\includegraphics[trim=5 5 5 5,clip,width=1.0\textwidth]{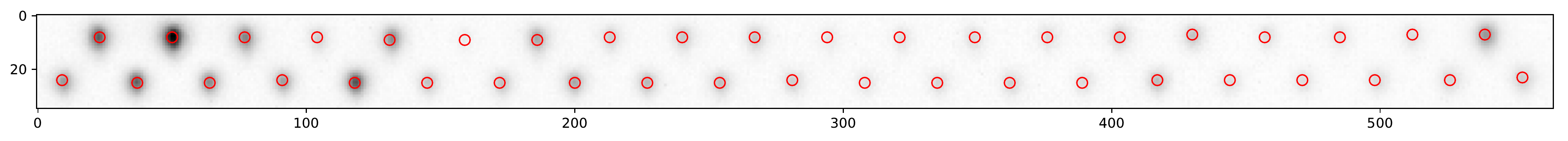}}

\caption{Top shows the output of the 6T component imaged on the infrared camera. The position of the outputs was derived from the image using the  OpenCV library. The numbers correspond to the outputs as defined in Figure~\ref{JH-testbench}. Bottom shows the physical size of the output channels used for extracting photometry and fringe data used in building the V2PM matrix.}
\label{DBCout}
\end{figure}

The inputs of the component are automatically found by performing a raster scan on the inputs using the stepper motors and measuring the integrated intensity from the outputs of the DBC. A XY map of the intensity is then built and the outputs found. We use the open--source library OpenCV to identify the inputs position and derive the XY coordinates\cite{Opencv_library}. Once we have the positions of the inputs for the two beams we can build the V2PM matrix of the component as described in\cite{2013ApOpt..52.4556S}. 

One beam only is injected and all the outputs recorded by the camera and the images saved as FITS files. The procedure is repeated for all six beams. This gives us a measurement of the photometry of the inputs. Two beams are then injected and a USB--controlled delay line scanned with steps that are 1/10 of the laser wavelength in use. An image of the camera is acquired for each step and recorded as a FITS image cube of the output of the component. The pixels corresponding to the output channels of the combiner are extracted and the photometry, obtained from single beams injection, subtracted from each channel separately. 

Figure~\ref{fringes} shows a time series from the channels extracted from output 16 of the waveguide for all possible combinations of two input beams (or baselines, on a real instrument). We calculate a FFT for each time series. The parameters extracted with the FFT are used as initial parameters for a sine plus cosine quadratures model. We extract the real and imaginary part of the interfering fields by applying a least--square fit to the fringe data using Equation~\ref{eq:fringe}:

\begin{equation}\label{eq:fringe}
f(t) = a  \cos(\omega t) + b  \sin(\omega t)
\end{equation}

The extracted parameters a and b are the real and imaginary part of the interference fringe. These values are placed in the V2PM matrix together with the photometry as described in \cite{2013ApOpt..52.4556S}. After the V2PM matrix is filled up it can be inverted to give the pixel to visibility matrix (P2VM) that relates the output recorded from the camera to the complex visibilities for all baselines and photometry for all telescopes in the array.  

\section{Conclusions}
We have presented the progress in the development of a six-inputs, J-band interferometric beam combiner based on the discrete beam combiner (DBC) concept. 

Before manufacturing the component we performed extensive numerical simulations with a coupled modes model of the DBC to identify the best input configuration and array length. The 41 waveguides were arranged on a zig-zag array that allows a simple optical setup for dispersing the light at the output of the waveguides.

The component we are currently developing is manufactured in borosilicate glass using the technique of multi-pass ultrafast laser inscription (ULI), using a mode-locked Yb:KYW laser at the wavelength of 1030 nm, pulse duration of 300 fs and repetition rate of 1 MHz. After annealing, the written components showed a propagation loss  less than 0.3 dB/cm and a negligible birefringence at a wavelength of 1310 nm, which makes the components suitable for un-polarized light operation. A single mode fiber-to-component insertion loss of 0.9 dB was measured. Work is currently in progress to characterize the components in spectro-interferometric mode with white light covering the J-band spectrum.

%%%%%%%%%%%%%%%%%%%%%%%%%%%%%%%%%%%%%%%%%%%%%%%%%%%%%%%%%%%%%
\acknowledgments{The author gratefully acknowledge the financial support from the Federal Ministry of Education and Research (BMBF), grant number 03Z22AN11. This research has made use of NASA's Astrophysics Data System Bibliographic Services. This research made use of the IPython package \cite{PER-GRA:2007}; matplotlib, a Python library for publication quality graphics \cite{Hunter:2007}.
}

%%%%%%%%%%%%%%%%%%%%%%%%%%%%%%%%%%%%%%%%%%%%%%%%%%%%%%%%%%%%%
%%%%% References %%%%%

\bibliographystyle{spiebib}
\bibliography{detectors,interf,textbooks,astro,imageproc,photonics}
\end{document}